\documentclass{optica-article}

\journal{opticajournal} 

\articletype{Research Article}

\usepackage{lineno}
\usepackage{comment}

\begin{document}

\title{Recording dynamic facial micro-expressions with a multi-focus camera array}

\author{Lucas Kreiss\authormark{1*,**},
Weiheng Tang\authormark{1*},
Ramana Balla\authormark{1}, 
Xi Yang\authormark{1}, 
Amey Chaware\authormark{1},
Kanghyun Kim\authormark{1}, 
Clare B. Cook\authormark{1}, 
Aurelien Begue\authormark{2}, 
Clay Dugo\authormark{2},
Mark Harfouche\authormark{2}, 
Kevin C. Zhou\authormark{1} and
Roarke Horstmeyer\authormark{1,2} }

\address{\authormark{1}Department of Biomedical Engineering,
Duke University, Durham, NC 27708 USA\\
\authormark{2}Ramona Optics Inc., 1000 W Main St., Durham, North Carolina 27701, USA\\
\authormark{*}equally contributing first authors}

\email{\authormark{**}lucas.kreiss@duke.edu} 


\begin{abstract*} 
We present an approach of utilizing a multi-camera array system for capturing dynamic high-resolution videos of the human face, with improved imaging performance as compared to traditional single-camera configurations. Employing an array of 54 individual high-resolution cameras, each with its own 13 megapixel sensor (709 megapixels total), we uniquely focus each camera to a different plane across the curved surface of the human face in order to capture dynamic facial expressions. Post-processing methods then stitch together each synchronized set of 54 images into a composite video frame. Our multi-focus strategy overcomes the resolution and depth-of-field (DOF) limitations for capturing macroscopically curved surfaces such as the human face, while maintaining high lateral resolution. Specifically we demonstrate how our setup achieves a generally uniform lateral resolution of 26.75 $\pm$ 8.8~\textmu m across a composite DOF of $\sim$43~mm that covers the entire face (85 cm$^2$+ FOV). Compared to a single-focus configuration this is almost a 10-fold increase in effective DOF. We believe that our new approach for multi-focus camera array video sets the stage for future video capture of a variety of dynamic and macroscopically curved surfaces at microscopic resolution.

\end{abstract*}

\section{Introduction}
Capturing high quality images and video of the human face, including its dynamic facial expressions, is crucial for applications in medical diagnostics, plastic surgery prognosis, and advanced facial recognition systems. Within this wide range of applications, recent example use cases include emotion detection~\cite{wang2022systematic,ko2018brief}, disease diagnosis~\cite{tu2019quantitative}, assessment of surgical outcomes~\cite{ccoban2021three,wang2021effect}, face prosthetic fabrication~\cite{buyukccavus2020evaluation} and even to detect genetic effects on specific facial features~\cite{crouch2018genetics}. 

While the majority of prior work in this space has focused on low-resolution imaging with a single camera, with images exhibiting up to several hundred \textmu m in optical resolution, there are increasing needs for high-resolution imagery of the human face. Resolution is especially important for detecting facial micro-expressions~\cite{wang2022systematic,davison2016samm}. For example, a 2014 database of 'high-resolution' dynamic facial expressions consisted of 2D texture videos with only 1,040~×~1,392~pixels/frame that were captured by 3 cameras at a rate of 25~fps~\cite{zhang2014bp4d}, which equates to approximately 400 \textmu m optical resolution across a standard face. A more recent review article on databases for emotion models, which analyzed 19 review papers and 350+ researches papers~\cite{wang2022systematic}, found that the database with the highest resolution was the Spontaneous Micro-Facial Movement (SAMM) data set~\cite{davison2016samm}, which consists of images with 2,048×1,088 pixels.

\begin{figure}[h!]
\centering\includegraphics[width=0.83\linewidth]{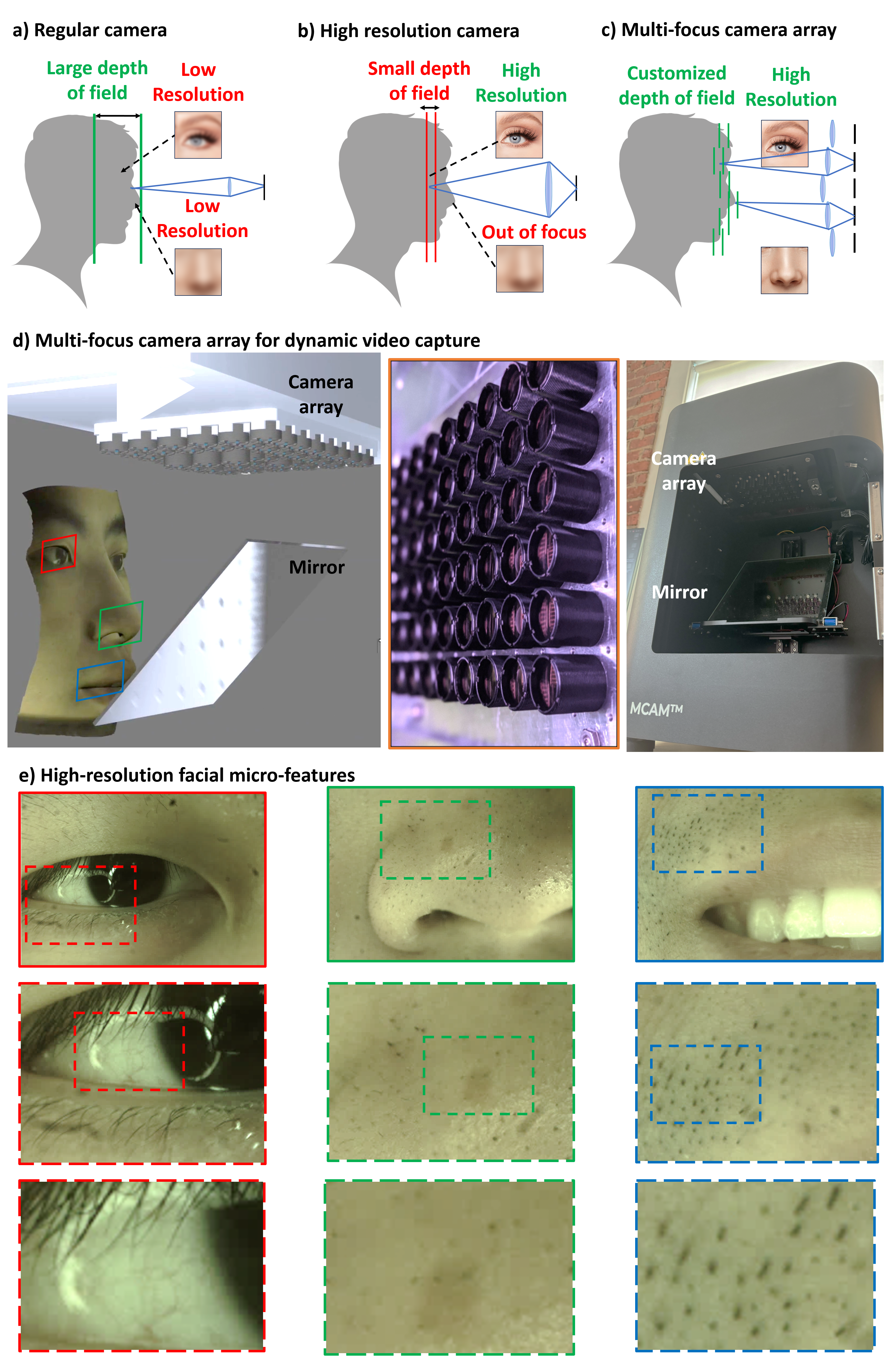}
\caption{Face imaging with the multi-camera array microscope (MCAM). (a) Regular cameras can image the large FOV of the face, albeit at low resolution. (b) Optics for high-resolution imaging, on the other hand, are limited by their depth of field (DOF) and can only bring a specific plane into focus. (c) A multi-focus camera array can enable high resolution imaging across a large FOV and at large effective DOF. (d) This concept can easily be applied by the MCAM, where each imaging unit in the array is adjusted to a different focal plane. These all-in-focus images allow to render digital avatars as well as (e) High resolution imaging of facial micro-features. }
\label{Fig1}
\end{figure}

A key challenge that prevents high-resolution video capture of the entire face is its highly curved nature. Use of a single camera presents the fundamental trade-off between the achievable lateral resolution, field of view (FOV), and depth of field (DOF) of its associated images. As highlighted in Fig.~\ref{Fig1}a, an everyday photographic camera can be configured, for instance, to image human faces with a sufficiently high DOF, meaning it can focus on a wider range of depth within the scene. However, its resolution is limited, making it inadequate for capturing fine details. In contrast, one might adopt a high-resolution camera (Fig.~\ref{Fig1}b) that can capture more intricate spatial details, but it comes with the cost of a reduced DOF, making it difficult or impossible to focus on features at varying distances simultaneously.

To overcome this tradeoff, we propose here the use of multiple, compact, closely spaced cameras that are designed to synchronously capture high-resolution video across different segments of the human face. By ensuring that each camera is focused appropriately on its corresponding face segment, for example via an auto-focus procedure, the images from all cameras can be stitched together to create a composite image that composes each final video frame (see Fig.~\ref{Fig1}c). While approaches using several uniquely angled cameras have been adopted in the past to capture en-face and highly oblique images~\cite{fyffe2014driving}, little work has pushed such a technology to the use of dozens or many dozens of cameras in an attempt to approach extremely high (potentially cellular-scale) resolution, which is a key focus here.

In this work, we demonstrate the use of 54 cameras positioned in a compact 9~$\times$~6 array to capture dynamic videos of the human face. By ensuring each camera is appropriately focused at its corresponding facial area, we demonstrate the ability to form in-focus projections that contain more than 13,000x9,000 pixels per frame across an approximate 10$\times$14 cm FOV and around 40~mm of depth profile, while still resolving features at only $\sim$25~\textmu m resolution. Compared to the SAMM data set~\cite{davison2016samm}, this represents more than a 50-fold increase in the total number of resolved pixels per video frame of the face, which we hypothesize can improve the information content within the various clinical, security, and entertainment applications that regularly rely on facial video capture noted above.

\section{Related Work}
Outside of the field of facial video capture, prior work has considered how to design imaging systems that can capture rapid dynamics of fine details across large curved surfaces. For example, to image neural cellular activity across the curved surface of the mouse cortex, Xie et al. recently employed multifocal fluorescence imaging of arbitrary surfaces (MFIAS) - a system using selective illumination and focal modulation by a digital micro-mirror device~(DMD) and a spinning disc with varying thickness to enable video-rate imaging at a frame rate of 10~fps, a lateral resolution of 7.4~\textmu m, a FOV of 7~mm and an effective DOF of several 100~\textmu m~\cite{xie2024multifocal}. The system was successfully used to capture calcium signaling of single neuron across the cortical surface of living mice~\cite{xie2024multifocal}. Alternatively, extended depth of field (EDOF) methods, which use point-spread function engineering to create and then digitally remove depth-dependent blurring, have been developed to enlarge the depth of field of cameras~\cite{dowski1995extended,teng2024hybrid} and microscopes~\cite{meng2022drop,ram2023caustic}. While the above approaches may enhance the ability to image main features in sharp focus across highly curved surfaces, they do not directly increase the native spatial resolution of the single-lens, single-sensor camera across a specific FOV. 

To jointly capture high resolution and large FOV images, both high optical resolution and a large number of pixels are needed. To achieve this goal, prior work has used camera arrays, which consist of multiple smaller individual high-resolution cameras and microscopes that jointly image unique areas of a larger FOV in parallel~\cite{thomson2022gigapixel,harfouche2023imaging,zhou2023parallelized,yang2023multi,zhou2024computational,Kim2024}. By stitching composite images from all cameras in the array, one can create an enlarged effective system FOV. For example, a 5~$\times$~7~array was used in the real-time ultra-large-scale high-resolution (RUSH) microscope~\cite{fan2019video}, a 3~$\times$~2~array was used in Barlow et al.~\cite{barlow2022megapixel}, and a 4~$\times$~4~array was used in Cibir et al.~\cite{cibir2023complexeye}. Recently, multi-camera array microscopes (MCAMs) ~\cite{harfouche2023imaging,yang2023multi,Kim2024} have been developed for large FOV, high-resolution capture with up to 96 cameras~\cite{thomson2022gigapixel}, which have also been applied to axial scanning of centimeter scaled objects~\cite{zhou2024computational} and even high-speed 3D reconstruction of freely moving organisms, like harvester ants and fruit flies~\cite{zhou2023parallelized}. In this work, we take advantage of the ability to uniquely focus each compact camera unit, while simultaneously maintaining sufficient spatial overlap between adjacent imaging FOV's, to accurately stitch together dozens of images into all-in-focus composite video frames of the face's highly curved surface. This novel approach addresses the joint challenge of capturing video with an extended depth of field, high-resolution and large FOV that is required to accurately map the dynamic morphology of the human face at resolutions that approach the cellular scale.  

\section{Methods}

As shown in Fig.~\ref{Fig1} and supplementary video 1, our face imaging system is based on a multi-camera array microscope (MCAM) which consists of 54 paralleled cameras in a 9~$\times$~6 grid. First, the focal planes of each individual camera in this MCAM system were adjusted to enable multi-focus imaging, as described below. Secondly, we characterized the imaging performance in terms of focal plane, resolution, DOF of each camera, and EDOF of the entire array. Finally, we employed this system for high-resolution image and video capture of human faces.  

\subsection{System configuration for multi-focus imaging}

\begin{figure}[t!]
\centering\includegraphics[width=0.75\linewidth]{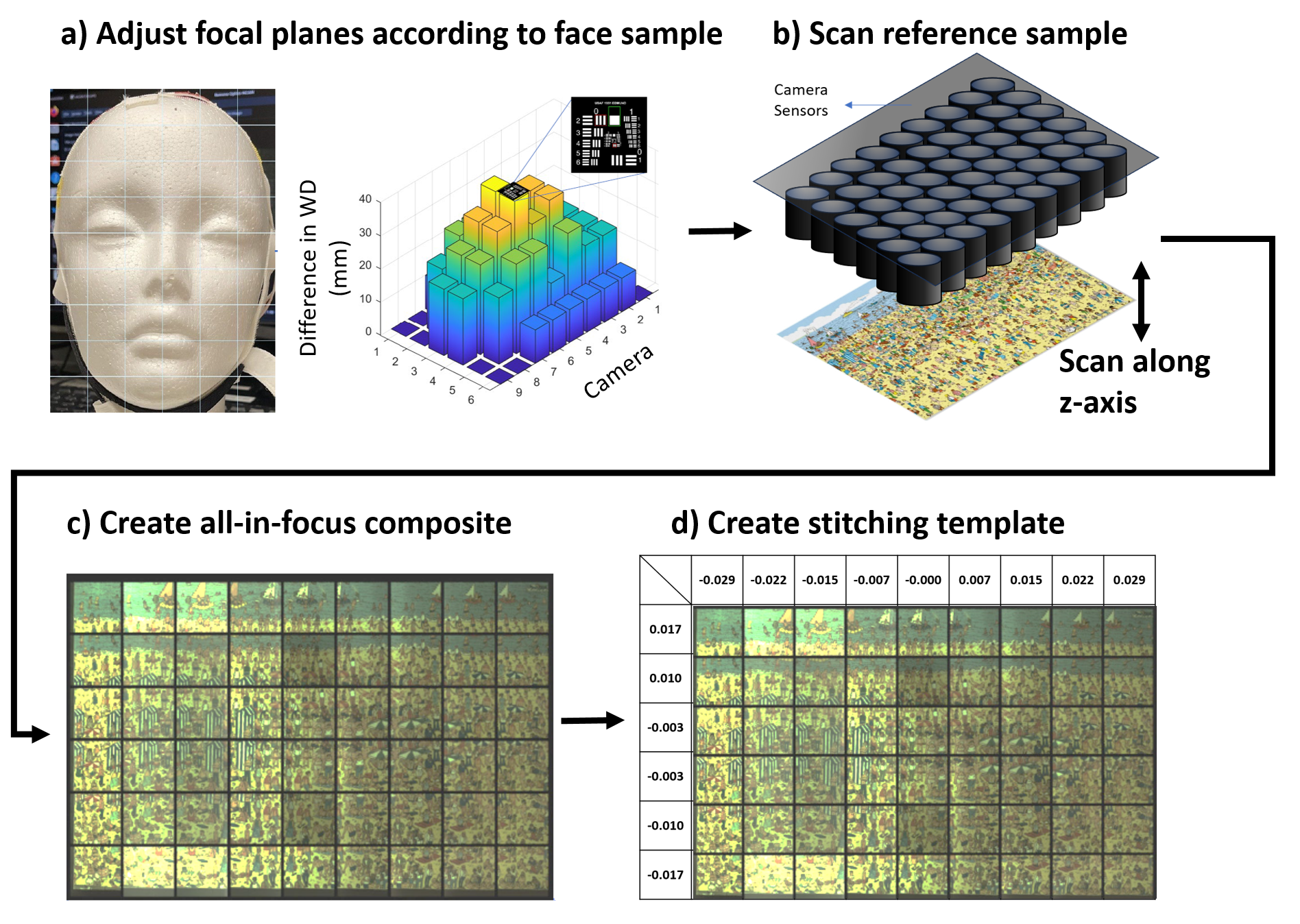}
\caption{Configuration of the multi-focus MCAM. (a) The object distance was adjusted for each camera according to a styrofoam face model. (b) a reference sample with a high number of spatial features was imaged to calibrate the stitching algorithm to obtain an (c) all-in-focus composite image, as well as (d) a stitching template.}
\label{Fig2}
\end{figure}

Each camera in the array consists of a lens with a focal length (f) of 25.05~mm, and an object side numerical aperture (NA) of 0.04, as well as a 13 megapixel CMOS sensor (ONSemi AR1335, 3,120×4,208 pixels, $\delta$=1.1~\textmu m pixel width)~\cite{harfouche2023imaging}. The cameras were tiled at a pitch of 13.5~mm on a single PCB board~\cite{harfouche2023imaging}. The lens array of the MCAM is fixed in a downward-facing orientation and the default object distance is 240~mm~\cite{harfouche2023imaging}. 
 
We utilized an anatomically accurate styrofoam face model to estimate a reference facial depth at each camera position using a digital vernier scale. This facial depth profile ranges between 0~-~40~mm, as seen in  Fig.~\ref{Fig2}a. To match these depth values for each camera, a flat, feature-rich reference sample was moved to these aimed axial positions and the lenses were axially adjusted, until the resolution target was in focus (see Fig.~\ref{Fig2}b). The maximal difference in focal positions between directly adjacent cameras was 20~mm. This resulted in cameras exhibiting object distances ($d_o$) that ranged between 200-240~mm, with corresponding image distances that range between 27.9~mm and 28.5~mm, calculated using $d_i = \frac{1}{1/f - 1/d_o}$. Accordingly, an auto-focusing module that could be deployed with the utilized sensor array would need to adjust lens positions a total distance of 0.6~mm.

Imaging with this multi-focus camera array allows to generate all-in-focus images by stitching the 54 individual images from each camera, using the Hugin stitching algorithm~\cite{hugin}. The parameters for that algorithm were calibrated initially on the in-focus image of the reference sample of each camera (see Fig.~\ref{Fig2}c). These parameters were then saved and used for subsequent face stitching (Fig.~\ref{Fig2}d). The raw images of each individual camera were recorded at 3,072x3,072 pixels, and the final, stitched multi-focus image resulted in 13,394x9,062 pixels of 112x75~mm².


\begin{figure}[t]
\centering\includegraphics[width=0.9\linewidth]{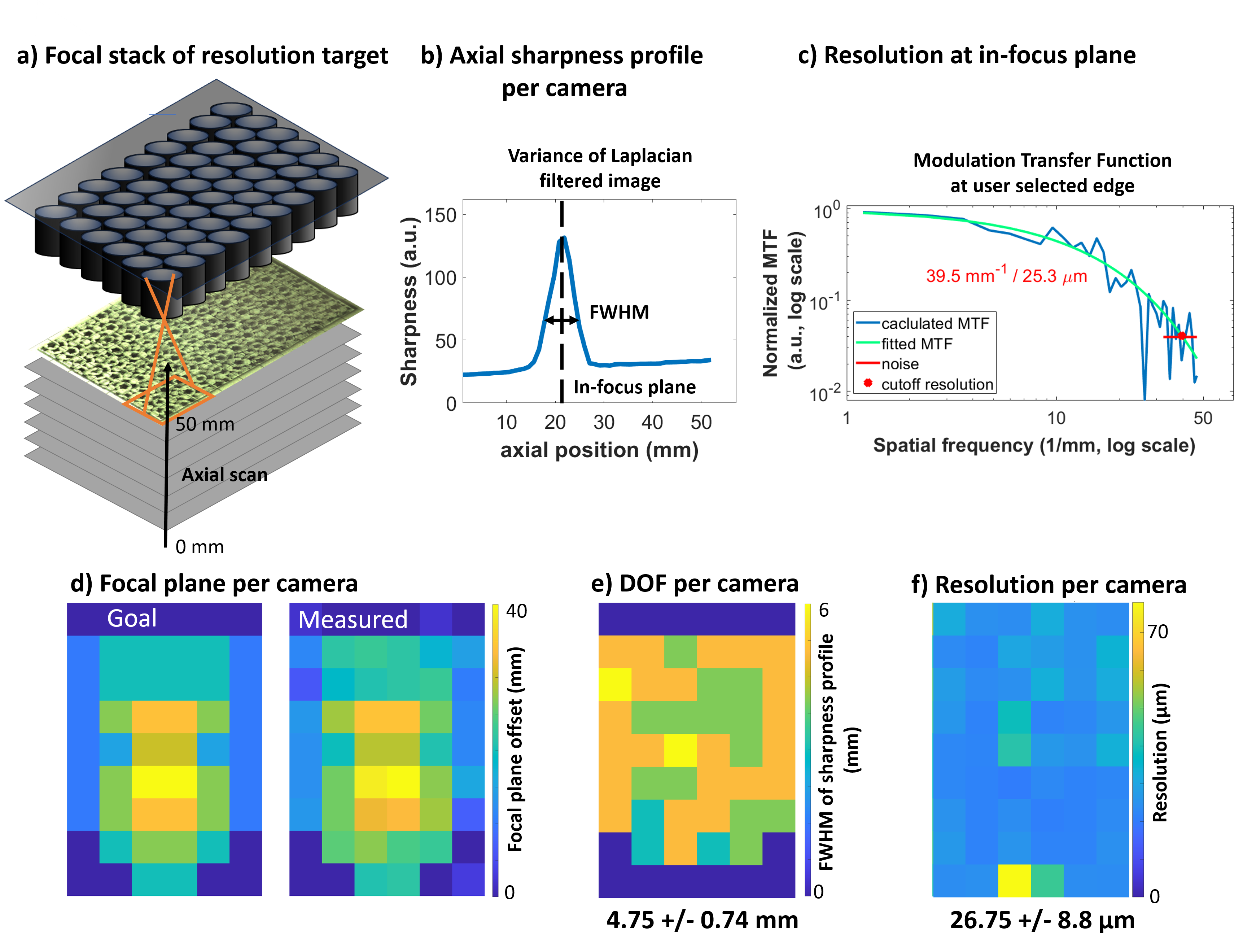}
\caption{Imaging performance (a) A focal stack of a resolution target was scanned at 1~mm step size within a range of 0-50mm. (b) For each camera and at each axial position, the sharpness was calculated to obtain an axial sharpness profile and identify the most in-focus plane. The FWHM was used as depth of field (DOF) estimation. (c) Additionally, a sharp edge in the most in-focus image was selected to obtain a resolution measurement from the modulation transfer function (MTF). Finally,  we plotted focal plane (d), DOF (e) and resolution (f) for each camera in the array.}
\label{Fig3}
\end{figure}

\subsection{Characterization of the multi-focus imaging system}

\hspace{7pt}
In that configuration, adjacent cameras had a $\sim$60~\% overlap in their individual FOV, and slightly different magnifications and FOV, due to the different object distances of each camera. The exact magnification M was calculated from $d_o$ and the focal distance of the lens (f) as:
\begin{equation}
    M=\frac{f}{d_{o}-f}
\end{equation}

Accordingly, the diameter of the FOV of each camera ranges between 23 and 29~mm. To test how differences in the height profile of a sample affect images, we calculated the parallax shift ($\Delta x$) that can be caused from certain height deviation ($\Delta h$) within each individual camera's FOV. Depending on the position of the height deviation within the FOV (r) and the camera-specific object distance $d_o$, we can calculate the chief ray angle (CRA) as: 
\begin{equation}
CRA(r) = arctan( \frac{r}{d_o} )
\end{equation}

Which leads to a parallax shift $\Delta x$ of
\begin{equation}
\Delta x = sin(CRA) \cdot \Delta h
\end{equation}

We tested the worst case, where the height deviation will be located at the very edge of a given camera's FOV and an intermediate case, where it is located at the mid-point between center and outer edge of the FOV. The parallax shift from a resolution-limited spot in the center of the FOV is negligibly small. We then calculated the lateral resolution, as explained below, and compared it to this parallax shift caused by height differences in the sample.





In order to quantify the exact focal plane, lateral resolution, DOF of each camera, and extended DOF of the entire array, we used a translation stage (Thorlabs XR25 Series) with a precision of 0.01~mm to take an axial stack of a custom-designed resolution target (see Fig.~\ref{Fig3}a). Therefore, an image was captured by the entire camera array at every 1~mm within a range of -1 to 50~mm from the default working distance of 240~mm (see '0' in the focal map of Fig.~\ref{Fig1}). 

Image sharpness was quantified at each plane and for each camera, by convolving the image with a Laplacian filter and calculating the total variance of the filtered image as a measure for the overall sharpness~\cite{pertuz2013analysis}.

\begin{equation}
    F_{laplacian} = \begin{bmatrix}
1 & 1 & 1\\
1 & -8 & 1 \\
1 & 1 & 1\\
\end{bmatrix}
\end{equation}

This sharpness metric was then plotted against the axial position in the focal stack (see Fig.~\ref{Fig3}b) and the plane of maximum sharpness was determined as in-focus plane for that camera at the respective object distance $d_o$. The full width at half maximum (FWHM) of this axial sharpness profile was used as a metric for the DOF of the respective camera. 

To quantify the lateral resolution, a sharp, vertical edge at the in-focus image plan  of each camera was manually selected (see Fig.~\ref{Fig3}c). This region of interest (ROI) was used to compute the edge spread function (ESF) as the mean gray scale intensity across the edge. The line spread function (LSF) was then calculated as the derivative of the ESF, which in term served as base for calculating the optical transfer function (OTF) by taking the fast Fourier transform (FFT) of the LSF. Finally, the absolute value of the OTF is used as modulation transfer function (MTF), which was plotted against the spatial frequency between 0 and 50 1/mm. The MTF was split into two regions: the noise cutoff at high spatial frequencies and the exponential decay for the MTF region before that. The latter regions was fitted by an exponential function and the intersection between MTF function and the mean noise was determined as the cutoff spatial frequency. The inverse of this cutoff spatial frequency was used as definition of the lateral resolution for each camera. 

\subsection{Face imaging with the multi-focus imaging system}

Upon confirming the correct configuration of the multi-focus camera array and quantifying its imaging performance, we tested our system on imaging the natural curvature of real human faces. For that purpose, a 20~x~20~cm mirror was placed at 45-degree in the optical path to enable direct face imaging (see Fig.\ref{Fig1}d). The subjects placed their head onto a stationary chin rest to ensure a comfortable position. The faces were illuminated with three ring lights (Qasim White 110MM 81SMD COB LED Halo Ring), one placed directly in front of the face but behind the mirror, and one from the left and right side of the face, while image and video data were captured by the multi-focus MCAM system. 



\begin{figure*}[ht!]
\centering\includegraphics[width=\linewidth]{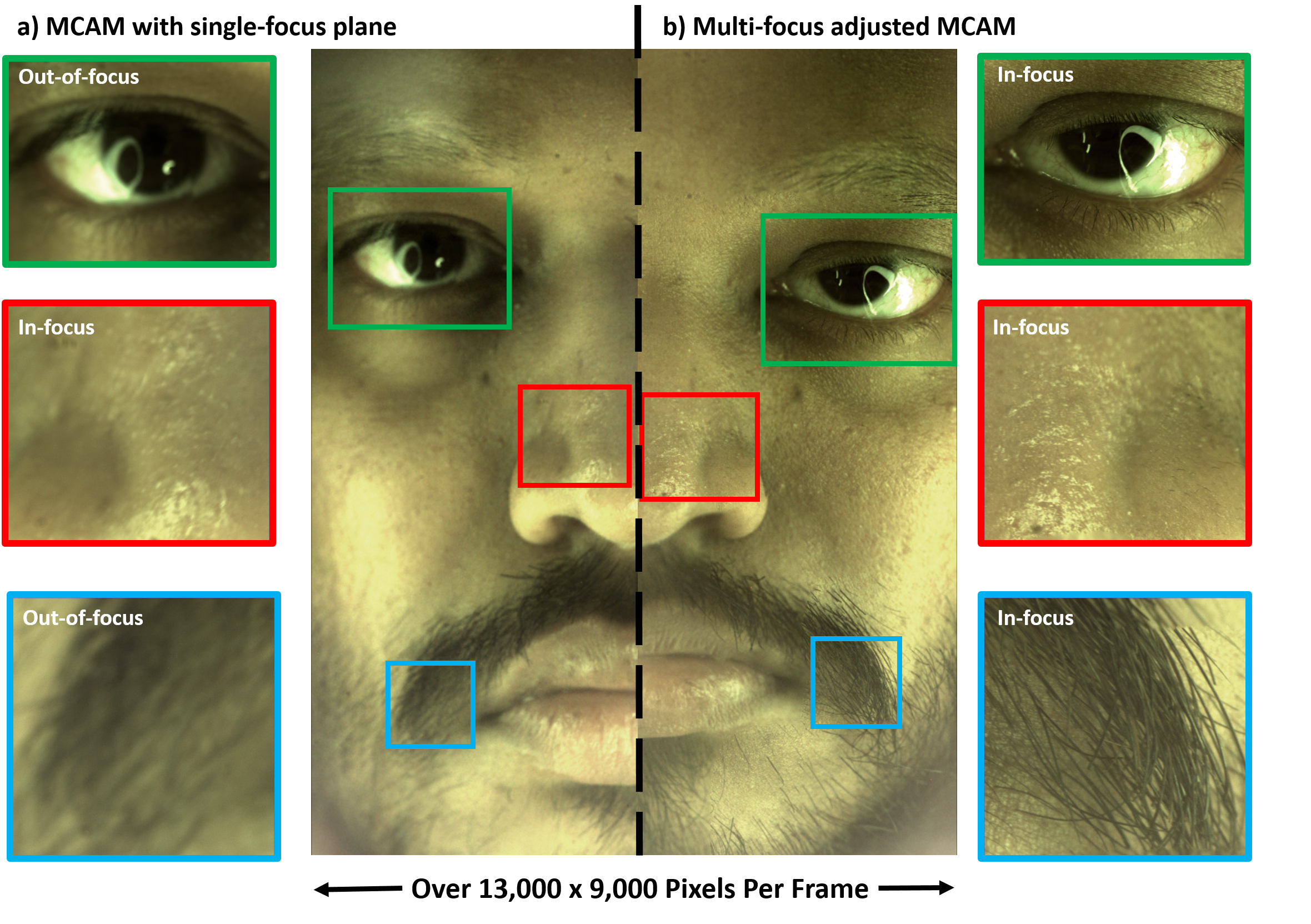}
\caption{High-resolution, micro-feature face imaging over a large FOV and a large range of DOF. (a) MCAM in a conventional configuration with all cameras focused at the same plane, where certain facial features are out-of-focus. (b) Re-focused MCAM configuration where all facial features from different depths are in focus.}
\label{Fig4}
\end{figure*}

\section{Results}
\subsection{Extended DOF, lateral resolution and parallax shift of the multi-focus imaging system}

The magnification varied between 0.1165 and 0.1463 across the entire array according to the varying object distances of each camera, which was taken into account to calculate the image pixel size during the resolution quantification (Fig.~\ref{Fig3}c). The maximal difference in magnification between directly adjacent cameras $\Delta M_{max}$ was 0.0153 or $\sim$ 10~\%. 

The measurement of the focal plane of each individual camera in the array is in very good agreement with the aimed multi-focus depth profile, as shown in Fig.~\ref{Fig3}d. The full range of measured focal planes was 43.64~mm, which defines the extended DOF of the entire multi-focus array. The DOF of each individual camera was measured as 4.7 $\pm$ 0.7~mm, and was distributed uniformly across all cameras (see Fig.~\ref{Fig3}e). Thus, our multi-focus configuration results in an almost 10-fold increase in EDOF compared to the DOF of a single-focus configuration. The slight difference of a few mm between goal and measured depth map (see Fig.~\ref{Fig3}d) is within the acceptable margin for error, according to the individual DOF of each camera (see Fig.~\ref{Fig3}e). In a similar fashion, the lateral resolution at the most in-focus-plane of each camera was measured as 26.75 $\pm$ 8.8~\textmu m (see Fig.~\ref{Fig3}e). This quantified resolution is in good agreement with the previously reported value of $\sim$20~\textmu m for a MCAM configuration, at a single focal plane, without the multi-focus setup~\cite{harfouche2023imaging}.

In the worst case for the parallax shift, when the height difference is at the extreme edge of a camera, a height between 350~\textmu m (at M=0.116) and 440~\textmu m (at M=0.146) would introduce a parallax shift similar to the measured lateral resolution. In the intermediate case, where it is at the mid point within the FOV, this height ranges between 650~\textmu m (M=0.116) and 830~\textmu m (M=0.146). Height difference above those values would start to introduce noticeable stitching errors.

\subsection{High-resolution imaging of micro-facial features}

In order to showcase the effect of our multi-focus imaging system, we compared face images from a MCAM system with uniform focal planes, i.e., all cameras focused at the same plane at WD=240~mm (see Fig.~\ref{Fig4}a) to images of the same face with our new, multi-focus MCAM (Fig.~\ref{Fig4}b). Due to the slight offset in magnification, discussed above, both face images in Fig.~\ref{Fig4}a~\&~b appear at slightly different image sizes. Nevertheless, these results demonstrate that the new system can capture facial micro-features at extremely high resolution, across the entire face and at different depths, while the conventional, single-focus MCAM suffers from the low DOF. 

\begin{figure*}[ht!]
\centering\includegraphics[width=\linewidth]{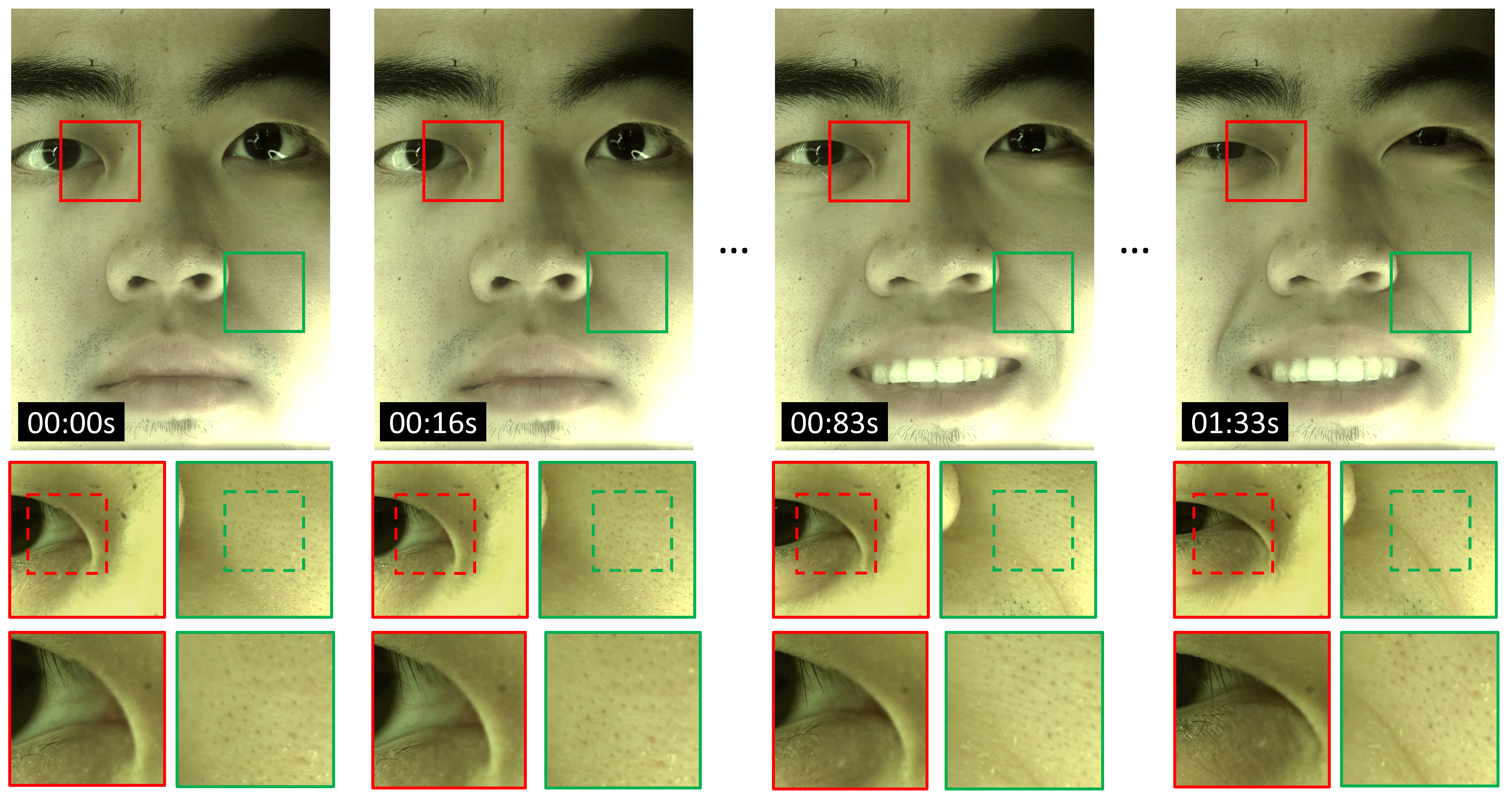}
\caption{Video frames from dynamic facial features, recorded with the MCAM at 12~fps, including with two selected zoom ins. See supplementary file 2 for the full video.}
\label{Fig5}
\end{figure*}

Furthermore, we recorded dynamic videos of facial movements at 12~fps, as shown in Fig.~\ref{Fig5} and supplementary video 2. These results show that our system is not only able to capture high-resolution micro features, but also to record video data of dynamically changing facial expression at high frame rate. As discussed by Harfouche et al.~\cite{harfouche2023imaging}, the frame rate of the MCAM is limited by the data transmission from the FPGA to the workstation and pixel binning could be used to reduce data size and increase frame rate, effectively trading-off pixel resolution against frame rate, if needed. 

We estimate the maximal facial height difference within a single camera's FOV (approximately 2.5x2.5~cm$^2$, depending on M) to be below 10~mm, e.g., for locations around the nose. According to the calculations above, this would result in a maximal parallax shift between 1~\textmu m (resolution-limited spot in the center of a single camera's FOV) and 600~\textmu m or 28 times the lateral resolution (when located at the far edge of a single camera's FOV and at M=0.116). The combined effect of this parallax shift and slightly different magnifications between adjacent cameras ($\sim$ 10~\%) can introduce stitching errors that manifest as slightly blurred lines or areas. As apparent from images of reference samples (see Fig.~\ref{Fig2}~\&~\ref{Fig3}) or human faces (see Fig.~\ref{Fig4}~\&~\ref{Fig5}), these errors are relatively small compared to the large FOV and high total number of resolved pixels.

\section{Conclusion}
In this study, we have demonstrated the significant advantages of using a multi-focus, multi-camera array system, like the MCAM, for recording dynamic facial expressions at high resolution. 

By adjusting the focal plane of each camera to focus on different regions on the face, we achieved a significantly extended depth of field (EDOF) of $\sim$43~mm for the entire array, across several cm$^2$ FOV, while maintaining a resolution of only a few tens of \textmu m. Since each single microscope unit has an individual DOF of $\sim$4.7~mm, our systems allows a certain deviation from the configuration profile, and thus the setup is compatible with several different face geometries, without re-calibration. Compared to the MFIAS system~\cite{xie2024multifocal}, our multi-focus MCAM has a similar frame rate, a slightly reduced lateral resolution, but a greatly increased FOV and DOF.

We showed that this approach successfully addresses the limitations of single-camera systems, particularly in balancing high resolution and DOF. This capability allows the capture of intricate facial details, while maintaining an effective DOF that covers the whole face depth profile. The versatility and adaptability of our system to different facial profiles further enhance its applicability in diverse real-world scenarios. We believe that this technological advancement is crucial for applications requiring precise and high-quality facial imaging, such as medical diagnostics, plastic surgery prognosis, and advanced facial recognition systems. As shown in our results of generating digital avatars (Fig.~\ref{Fig1}d), we believe this system further has a great potential for applications in augmented or virtual reality. 

Future work systems will integrate lenses with tunable focus to enable automated re-focusing of each camera. Additionally, we foresee integration of advanced algorithms for automatic calibration and real-time processing to further improve the performance and usability of the MCAM system. 

\section{Supplementary materials}

\subsection{Supplementary video 1}
A video showing the principle of using a multi-focus MCAM for face imaging.
\subsection{Supplementary video 2}
A video showing dynamic facial expression at great resolution (recorded at 12~fps, played at 3~fps).

\section{Back matter}

\begin{backmatter}
\bmsection{Funding}
Research reported in this publication was supported by the Office of Research Infrastructure Programs (ORIP), Office Of The Director, National Institutes Of Health of the National Institutes Of Health and the National Institute Of Environmental Health Sciences (NIEHS) of the National Institutes of Health under Award Number R44OD024879, the National Cancer Institute (NCI) of the National Institutes of Health under Award Number R44CA250877, the National Institute of Biomedical Imaging and Bioengineering (NIBIB) of the National Institutes of Health under Award Number R43EB030979,  the National Science Foundation under Award Number 2036439, and a Duke-Coulter Translational Partnership Grant. KCZ was supported in part by the Schmidt Science Fellows, in partnership with the Rhodes Trust.


\bmsection{Disclosures}
R.H. and M.H. are cofounders of Ramona Optics, Inc., which is commercializing multi-camera array microscopes. C.B.C., A.B., C.D. and M.H. are or were employed by Ramona Optics, Inc. during the course of this research. K.C.Z. is a consultant for Ramona Optics, Inc. The remaining authors declare no competing interests.

\bmsection{Data Availability Statement}
Data underlying the results presented in this paper are not publicly available at this time but may be obtained from the authors upon reasonable request.

\end{backmatter}

\bibliography{sample}

\begin{thebibliography}{10}
\newcommand{\enquote}[1]{``#1''}

\bibitem{wang2022systematic}
Y.~Wang, W.~Song, W.~Tao, \emph{et~al.}, \enquote{A systematic review on affective computing: Emotion models, databases, and recent advances,} {\protect\JournalTitle{Information Fusion}} \textbf{83}, 19--52 (2022).

\bibitem{ko2018brief}
B.~C. Ko, \enquote{A brief review of facial emotion recognition based on visual information,} {\protect\JournalTitle{sensors}} \textbf{18}, 401 (2018).

\bibitem{tu2019quantitative}
L.~Tu, A.~R. Porras, A.~Oh, \emph{et~al.}, \enquote{Quantitative evaluation of local head malformations from 3 dimensional photography: application to craniosynostosis,} in \emph{Medical Imaging 2019: Computer-Aided Diagnosis,}  vol. 10950 (SPIE, 2019), pp. 798--803.

\bibitem{ccoban2021three}
G.~{\c{C}}oban, {\.I}.~Yavuz, and A.~E. Demirba{\c{s}}, \enquote{Three-dimensional changes in the location of soft tissue landmarks following bimaxillary orthognathic surgery.} {\protect\JournalTitle{Journal of Orofacial Orthopedics/Fortschritte der Kieferorthopadie}} \textbf{82} (2021).

\bibitem{wang2021effect}
P.-F. Wang, D.~C. Pascasio, S.~H. Kwon, \emph{et~al.}, \enquote{The effect of absorbable and non-absorbable sutures on nasal width following cinch sutures in orthognathic surgery,} {\protect\JournalTitle{Symmetry}} \textbf{13}, 1495 (2021).

\bibitem{buyukccavus2020evaluation}
M.~H. B{\"u}y{\"u}k{\c{c}}avus, Y.~Findik, and T.~Baykul, \enquote{Evaluation of changes in nasal projection after surgically assisted rapid maxillary expansion with 3dmd face system,} {\protect\JournalTitle{Journal of Craniofacial Surgery}} \textbf{31}, e462--e465 (2020).

\bibitem{crouch2018genetics}
D.~J. Crouch, B.~Winney, W.~P. Koppen, \emph{et~al.}, \enquote{Genetics of the human face: Identification of large-effect single gene variants,} {\protect\JournalTitle{Proceedings of the National Academy of Sciences}} \textbf{115}, E676--E685 (2018).

\bibitem{davison2016samm}
A.~K. Davison, C.~Lansley, N.~Costen, \emph{et~al.}, \enquote{Samm: A spontaneous micro-facial movement dataset,} {\protect\JournalTitle{IEEE transactions on affective computing}} \textbf{9}, 116--129 (2016).

\bibitem{zhang2014bp4d}
X.~Zhang, L.~Yin, J.~F. Cohn, \emph{et~al.}, \enquote{Bp4d-spontaneous: a high-resolution spontaneous 3d dynamic facial expression database,} {\protect\JournalTitle{Image and Vision Computing}} \textbf{32}, 692--706 (2014).

\bibitem{fyffe2014driving}
G.~Fyffe, A.~Jones, O.~Alexander, \emph{et~al.}, \enquote{Driving high-resolution facial scans with video performance capture,} {\protect\JournalTitle{ACM Transactions on Graphics (TOG)}} \textbf{34}, 1--14 (2014).

\bibitem{xie2024multifocal}
H.~Xie, X.~Han, G.~Xiao, \emph{et~al.}, \enquote{Multifocal fluorescence video-rate imaging of centimetre-wide arbitrarily shaped brain surfaces at micrometric resolution,} {\protect\JournalTitle{Nature Biomedical Engineering}} \textbf{8}, 740--753 (2024).

\bibitem{dowski1995extended}
E.~R. Dowski~Jr and W.~T. Cathey, \enquote{Extended depth of field through wave-front coding,} {\protect\JournalTitle{Applied optics}} \textbf{34}, 1859--1866 (1995).

\bibitem{teng2024hybrid}
M.~Teng, H.~Lou, Y.~Yang, \emph{et~al.}, \enquote{Hybrid all-in-focus imaging from neuromorphic focal stack,} {\protect\JournalTitle{IEEE Transactions on Pattern Analysis and Machine Intelligence}}  (2024).

\bibitem{meng2022drop}
Q.~Meng, Y.~Li, Y.~Yu, \emph{et~al.}, \enquote{A drop-in, focus-extending phase mask simplifies microscopic and microfluidic imaging systems for cost-effective point-of-care diagnostics,} {\protect\JournalTitle{Analytical Chemistry}} \textbf{94}, 11000--11007 (2022).

\bibitem{ram2023caustic}
R.~Ram, S.~Kumar, C.~Huang, \emph{et~al.}, \enquote{Caustic wavefront encoded imaging for snapshot three-dimensional fluorescence microscopy,} {\protect\JournalTitle{Research Square Preprint}}  (2023).

\bibitem{thomson2022gigapixel}
E.~E. Thomson, M.~Harfouche, K.~Kim, \emph{et~al.}, \enquote{Gigapixel imaging with a novel multi-camera array microscope,} {\protect\JournalTitle{Elife}} \textbf{11}, e74988 (2022).

\bibitem{harfouche2023imaging}
M.~Harfouche, K.~Kim, K.~C. Zhou, \emph{et~al.}, \enquote{Imaging across multiple spatial scales with the multi-camera array microscope,} {\protect\JournalTitle{Optica}} \textbf{10}, 471--480 (2023).

\bibitem{zhou2023parallelized}
K.~C. Zhou, M.~Harfouche, C.~L. Cooke, \emph{et~al.}, \enquote{Parallelized computational 3d video microscopy of freely moving organisms at multiple gigapixels per second,} {\protect\JournalTitle{Nature photonics}} \textbf{17}, 442--450 (2023).

\bibitem{yang2023multi}
X.~Yang, M.~Harfouche, K.~C. Zhou, \emph{et~al.}, \enquote{Multi-modal imaging using a cascaded microscope design,} {\protect\JournalTitle{Optics Letters}} \textbf{48}, 1658--1661 (2023).

\bibitem{zhou2024computational}
K.~C. Zhou, M.~Harfouche, M.~Zheng, \emph{et~al.}, \enquote{Computational 3d topographic microscopy from terabytes of data per sample,} {\protect\JournalTitle{Journal of Big Data}} \textbf{11}, 62 (2024).

\bibitem{Kim2024}
K.~Kim, A.~Chaware, C.~Cook, \emph{et~al.}, \enquote{Rapid 16-gigapixel imaging at cellular resolution for digital pathology with the multi-camera array scanner,} {\protect\JournalTitle{npj Imaging}} \textbf{-}, -- (2024).

\bibitem{fan2019video}
J.~Fan, J.~Suo, J.~Wu, \emph{et~al.}, \enquote{Video-rate imaging of biological dynamics at centimetre scale and micrometre resolution,} {\protect\JournalTitle{Nature Photonics}} \textbf{13}, 809--816 (2019).

\bibitem{barlow2022megapixel}
I.~L. Barlow, L.~Feriani, E.~Minga, \emph{et~al.}, \enquote{Megapixel camera arrays enable high-resolution animal tracking in multiwell plates,} {\protect\JournalTitle{Communications biology}} \textbf{5}, 253 (2022).

\bibitem{cibir2023complexeye}
Z.~Cibir, J.~Hassel, J.~Sonneck, \emph{et~al.}, \enquote{Complexeye: a multi-lens array microscope for high-throughput embedded immune cell migration analysis,} {\protect\JournalTitle{Nature Communications}} \textbf{14}, 8103 (2023).

\bibitem{hugin}
{Ramona Optics Inc.}, \enquote{{Ramona MCAM documentation},} \url{https://docs.ramonaoptics.com/python_module.html#owl-stitch-hugin-stitching}. {Hugin toolbox}.

\bibitem{pertuz2013analysis}
S.~Pertuz, D.~Puig, and M.~A. Garcia, \enquote{Analysis of focus measure operators for shape-from-focus,} {\protect\JournalTitle{Pattern Recognition}} \textbf{46}, 1415--1432 (2013).

\end{thebibliography}






\end{document}